\documentclass[12pt,preprint]{aastex}

\usepackage{natbib}
\usepackage{emulateapj5}

\shorttitle{Hydrodynamic Flows on Hot Jupiters}
\shortauthors{Langton \& Laughlin}

\begin{document}
\title{Observational Consequences of Hydrodynamic Flows on Hot Jupiters}
\author{Jonathan Langton\altaffilmark{1} and Gregory Laughlin\altaffilmark{2}}
\altaffiltext{1}{Physics Department, University of California at Santa Cruz}
\altaffiltext{2}{Astronomy and Astrophysics Department, University of California at Santa Cruz}
\begin{abstract}
We use a grid-based shallow water model to simulate the atmospheric dynamics of the transiting hot Jupiter HD 209458b.  Under the usual assumption that the planet is in synchronous rotation with zero obliquity, a steady state is reached with a well-localized cold spot centered $76^{\circ}$ east of the antistellar point.  This represents a departure from predictions made by previous simulations in the literature that used the shallow water formalism; we find that the disagreement is explained by the factor of 30 shorter radiative timescale used in our model.  We also examine the case that the planet is in Cassini state 2, in which the expected obliquity is $\sim 90^{\circ}$.  Under these circumstances, a periodic equilibrium is reached, with the temperature slightly leading the solar forcing.  Using these temperature distributions, we calculate disk-integrated bolometric infrared light curves from the planet.  The light curves for the two models are surprisingly similar, despite large differences in temperature patterns in the two cases.  In the zero-obliquity case, the intensity at the minimum is $66\%$ of the maximum intensity, with the minimum occuring $72^{\circ}$ ahead of transit.  In the high-obliquity case, the minimum occurs $54^{\circ}$ ahead of transit, with an intensity of $58\%$ of the maximum.
\end{abstract}


\section{Introduction}
Hot Jupiters orbit their parent stars with periods $P < 7$ days \citep{MQ95}.  They therefore receive a large stellar flux and  present an opportunity to study planetary behavior under conditions not seen in the solar system.  Furthermore, these planets are expected to have evolved tidally to spin-synchronicity.  In the event that the planetary spin and orbital angular momentum vectors are aligned, the planet keeps one hemisphere in perpetual daylight and the other in perpetual darkness.  The combination of intense irradiation and a permanent night side provides interesting modelling challenges that can be tested with observations in the infrared band \citep{Cha05,Dem05,Har06}.

The transiting extrasolar giant planet HD 209458b \citep{Cha00,Hen00} provides a particularly suitable opportunity to study these unique dynamics.  It orbits its primary at a distance of 0.045 AU with a period of 3.52474541 days \citep{Wit05}. The planet has a mass of $0.69 M_{J}$ and a radius of $1.32 R_{J} \pm 0.03 R_{J}$ \citep{Cha06}, $\sim 30\%$ larger than current planetary models predict \citep{Lau05}.  As reviewed by \citet{Cha06}, the planet must have a large source of interior heat; several explanations have been advanced to explain how this extra heat is produced.  In this Letter, we investigate the potential observational consequences of the hypothesis of \citet{WH05}, who suggest that HD 209458b may be trapped in Cassini state 2, in which the orbital angular momentum vector and the planetary spin vector precess at the same rate and the obliquity is forced to approach $90^{\circ}$ during planetary migration. In this scenario, the anomalous radius results from the heating caused by ongoing obliquity tides.  In this case, the atmospheric irradiation pattern is very different from that of the zero-obliquity case.  It is therefore interesting to investigate whether this difference results in a qualitatively different infrared light curve for the planet.

Several  hydrodynamic models have been used to investigate the atmosphere of HD 209458b under the assumption that the planet has zero obliquity \citep{Cho03, CS05, Bur05}.  \citet{CS05} model the atmosphere with a three-dimensional global circulation model.  They find that at atmospheric optical depth $\tau \sim 1$, a superrotating, supersonic equatorial jet advects the hottest region of the atmosphere $\sim 60^{\circ}$ downwind.  In contrast, \citet{Cho03} use a shallow water model with a radiative time constant $\tau_{rad} = 10$ days.  They obtain circumpolar vortices that sequester cold air and maintain a local temperature variation of up to 1000 K.  However, subsequent work \citep{Iro05} has suggested that the photospheric $\tau_{rad}$ should be on the order of hours, rather than days.  At present, there seems to be little agreement between models.  Fortunately, because the atmosphere is expected to exhibit extreme temperature variations, it is possible that observations of the planet's infrared signature will be able to resolve these discrepancies.  Therefore, a second goal in this Letter is to investigate the cause of the difference between our results and those of \citet{Cho03} and \citet{CS05}.

\section{Atmospheric Model}

Our simulations of the atmospheric flows are computed with a shallow water model similiar to that used in \citet{Cho03}.  This model  portrays the atmosphere as a thin layer of incompressible fluid at a constant $\rho$ and therefore is a reasonable approximation for subsonic stratified flows.  Framed in terms of the temperature $T$ and the horizontal wind velocity $\mathbf{v} = u \hat{\phi} + v \hat{\theta}$, the shallow water equations on an irradiated, rotating sphere are
\begin{eqnarray}
\frac{\partial T}{\partial t} &=& -\mathbf{v} \cdot \nabla T - T \nabla \cdot \mathbf{v} + F_{T}\\
\frac{\partial \mathbf{v}}{\partial t} &=& -\mathbf{v} \cdot \nabla \mathbf{v} - R \nabla T - f \hat{n} \times \mathbf{v},
\end{eqnarray}
where $R$ is the specific gas constant, $f = 2 \omega \sin \theta$ is the Coriolis parameter, $\hat{n}$ is the surface normal unit vector, and $F_{T}$ is the thermal forcing.  A rigorous treatment of the thermal forcing would involve a radiative transfer scheme; in this initial investigation, we choose instead to follow Cho's parametrization of the heating.  We assume an equilibrium temperature distribution 
\begin{equation}
\label{eqTeq}
T_{eq} = T_{eff} + \frac{\Delta T}{2} \cos \alpha,
\end{equation}
where $T_{eff}$ is the equilibrium blackbody temperature, $\Delta T$ is the maximum equilibrium day/night temperature difference at the layer being modeled in the absence of any hydrodynamic effects, and $\alpha$ is the angle between a given point $(\phi, \theta)$ on the planet's surface and the substellar point.  The equilibrium day/night temperature difference is a free parameter, and corresponds to the efficiency with which the planet can redistribute heat at an adiabatic depth.  This $\Delta T$ parameter should not be confused with the actual temperature variations that develop as the atmosphere evolves.  Rather, it measures the magnitude of the difference in forcing between the day and night hemispheres.  For the zero-obliquity case, we simply have $\cos \alpha = \cos \phi \cos \theta$.   We then impose Newtonian relaxation to this equilibrium temperature, so that
\begin{equation}
\label{eqFT}
F_{T} = -\frac{T - T_{eq}}{\tau_{rad}}.
\end{equation}
We adopt a radiative timescale $\tau_{rad} = $ 8 hours, the expected value at a pressure depth $p = 200$ mbar  \citep{Iro05}.
For the results presented here, we take $T_{eff} = 1400$ K, appropriate to a Bond albedo $A \sim 0.2$.  We further assume $\Delta T = 800$ K.  This relatively large value is motivated by the recent full-phase observations of $\upsilon$ Andromedae b by \citet{Har06}.

The shallow water equations are integrated using a vector spherical harmonic transform method proposed by \citet{AS99}.  The models are run at resolutions between $256 \times 127$ (T85) and $512 \times 257$ (T171).  Numerical stability is maintained using fourth-order hyperviscosity \citep{For98}, which prevents spectral blocking at high wavenumbers \citep{Boy00}.  We start with an isothermal ($T = T_{eff}$) atmosphere initially at rest, and integrate until either a steady state or a periodic equilibrium is reached, a process that is complete within 250 rotation periods at the radiative timescales considered.

\section{Results of the Zero-Obliquity Case}

\begin{figure}
\begin{center}
\includegraphics[clip = true, bb=0in 0in 8in 6in,trim = 0in 0in 0  0in, width=4in]{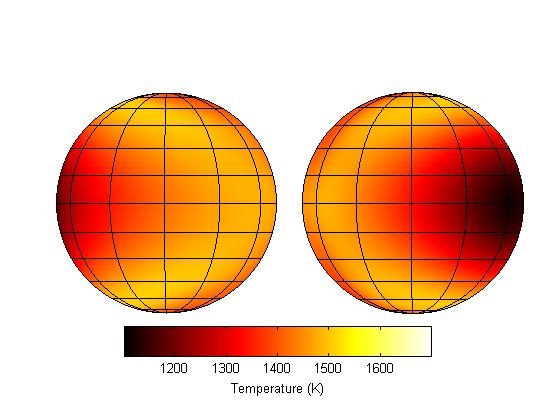}
\caption{Steady state temperature of HD 209458b in the zero-obliquity case.  Day side temperature distribution is shown on the left; night side on the right.}
\label{CS1fig}
\end{center}
\end{figure}

In the case that the planet has negligible obliquity, our model predicts that the atmosphere will exhibit an approximately steady state condition, as seen in Fig. \ref{CS1fig}.  The dominant feature is a localized cold spot on the night side.  However, this cold spot is not directly at the antistellar point; rather, its center is $76^{\circ}$ to the east; the coldest point is $63^{\circ}$ east of the antistellar point.  Interestingly, there appears to be a strong westward equatorial jet, with an average zonal wind speed of 780 m s$^{-1}$ at the equator.  The eastward displacement of the cold spot then results from the shallow water analog of the Bernoulli effect, rather than simple downwind advection of cold air.  The temperature distribution is quite asymmetric, with the maximum 110 K warmer than $T_{eff}$ and the minimum 280 K colder than $T_{eff}$.

This picture differs dramatically from the results of \citet{Cho03}, despite the similiarity of the model.  \citet{Cho03} start with random initial vorticity and use a long radiative time constant $\tau_{rad} = 10$ days.   After some tens of rotation periods, a dynamic equilibrium has evolved in which circumpolar vortices produce cold spots that orbit about the poles.  This leads to a situation where the temperature minimum can, at times, be located on the day side of the planet.

\begin{figure}
\begin{center}
\includegraphics[clip = true, bb=0in 0in 7in 11in,trim = 0in 0in 0  0in, width=4in]{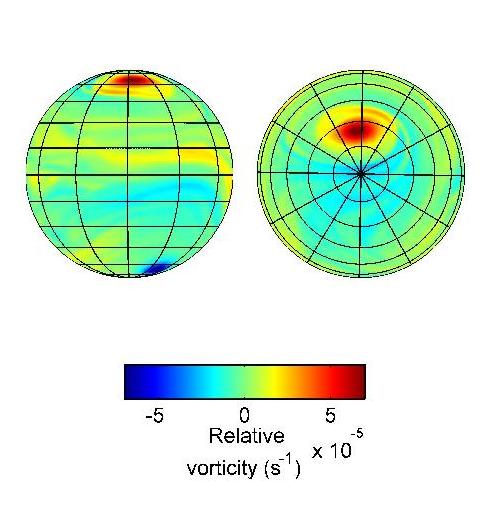}
\caption{Dynamic equilibrium reached from the evolution of vortical initial conditions in the absence of forcing.  Of note is the similarity to the results presented in \citet{Cho03}.}
\label{vortfig}
\end{center}
\end{figure}

While we start the atmosphere at rest, rather than using vortical initial conditions, the primary difference between our simulation and that of \citet{Cho03} is the radiative time constant; a value of $\tau_{rad} = 8$ hr is now believed to be more realistic \citep{Iro05}. Indeed, their $\tau_{rad}$ is sufficiently long that the forcing appears to have only negligible impact on the evolution of the atmosphere.  We ran simulations starting from vortical initial conditions as described by \citet{Cho03} with no thermal forcing;  within 40 rotation periods, the flow pattern was virtually identical to the Cho result \emph{with} forcing.  The end state of these simulations may be seen in Fig. \ref{vortfig}. The conclusion is that the atmospheric flow that \citet{Cho03} describe is more influenced by their initial conditions than it is by the intense stellar radiation.

While our results are more similar to those of \citet{CS05}, there are still some notable discrepancies.  \citet{CS05} use a fully three dimensional global circulation model, starting with a cold isothermal atmosphere at rest; after some time, a steady state emerges in the upper atmosphere.  They find that a strong eastward jet develops at the equator, advecting the hottest region $\sim 60^{\circ}$ downwind.  In contrast, we have a strong westward jet at the equator, with the cold spot being pulled upwind.  The reason for this difference is not immediately clear.  There is some indication that the development of superrotating jets requires three-dimensional processes \citep{CS05}.  If this is the case, they clearly would be unable to arise in our two-dimensional shallow water model.  On the other hand, a two-dimensional model allows higher horizontal resolution, which could reveal processes that would not be resolved in a full three-dimensional model.  Since the atmosphere at the infrared photosphere is expected to be radiative \citep{Iro05}, vertical mixing should be of secondary importance compared to the horizontal flow, and the use of a two-dimensional model is justified.

\section{Results of the High-Obliquity case}
 \Citet{WH05} suggest that the large radius of HD 209458b could be explained if the planet were in Cassini state 2.  A Cassini state is a spin-orbit resonance in which the planetary spin axis and the orbit normal precess about the same axis at the same rate.  In Cassini state 2, they are on opposite sides of the precession axis, and the planet must have non-zero obliquity.  It is expected that as the planet migrates inward and spin-synchronization occurs,  the axial tilt will increase until it is very close to $90^{\circ}$ \citep{WH05}.

In this case, the entire surface of the planet will receive stellar radiation; the time-dependent stellar elevation follows
\begin{equation}
\cos \alpha = \cos \theta \cos \theta_{*} \left(\cos \phi \cos \phi_{*} + \sin \phi \sin \phi_{*}  \right )+ \sin \theta \sin \theta_{*},
\end{equation}
where $\phi_{*} = -\omega t$ and $\theta_{*} = -\omega t$ give the position of the substellar point on the planetary surface.  The forcing itself is still given by equations $\ref{eqTeq}$ and $\ref{eqFT}$.
As may be expected, the atmospheric dynamics in this case differ greatly from those in the zero-obliquity case.  Here, a periodic equilibrium is reached, with one period per rotation.  The atmospheric response leads the forcing by $\sim 55^{\circ}$, so that the temperature begins to rise 13 hr before the onset of direct solar heating.  Temperature maxima are higher at the poles than at the equator as equatorial winds redistribute the heat.  Also notable are colder spots on the illuminated face produced by ring-shaped vortices that form at the poles and then break as they move to lower latitudes.  The temperature distribution is considerably more symmetric than that of the zero-obliquity case; the maximum and the minimum are $\sim 250 K$ warmer and colder than $T_{eff}$, respectively.

\section {Infrared Signatures}

Having determined the photospheric temperature, it is possible to calculate the relative intensity of the planet's infrared emissions.  We assume that the planet radiates as a blackbody; then the intensity of radiation emitted is 
\begin{equation}
I = \sum (\hat{n} \cdot \hat{r}) kT^{4} \cos \theta \Delta \theta \Delta \phi,
\end{equation}
where $\hat{n}$ is a unit surface normal vector, $\hat{r}$ is a unit vector directed from the planet towards earth, and $k$ is a normalization constant chosen so that $I = 1$ at maximum intensity.  We sum over all grid points on the portion of the planet facing earth (i.e., where $\hat{n} \cdot \hat{r}>0$).

\begin{figure}
\begin{center}
\includegraphics[clip = true, bb=0in 0in 5in 6in,trim = 0in 0in 0  0in, width=4in]{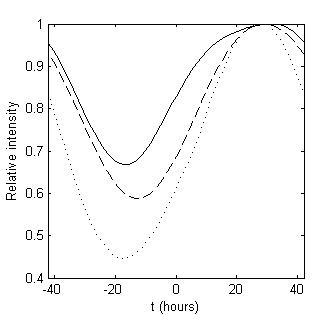}
\caption{Infrared emissions from HD 209458b over one full rotation period. Transit occurs at $t=0$. The solid line shows the light curve from the zero-obliquity case; the dashed line shows the light curve from the high-obliquity case. The light curve obtained by \citet{CS05}, shown by the dotted line, is included for comparison.}
\label{lcfig}
\end{center}
\end{figure}

The resulting light curves for both zero obliquity and high obliquity are seen in Fig. \ref{lcfig}.  Due to the temperature asymmetry apparent in the zero-obliquity case, the maxima are considerably wider than the minima.  This is the major contrast between the two cases; in the case of high obliquity, the maxima and minima are symmetric.  However, neither curve diverges significantly from a sinusoidal shape.  In both cases, the extrema lead the forcing.  In the zero-obliquity case, the minimum has a phase offset of $72^{\circ}$, while the maximum is offset by only $43^{\circ}$.  The high-obliquity case is closer to symmetric, with slightly smaller phase offsets of $54^{\circ}$ and $58^{\circ}$ at the minimum and maximum, respectively.

These light curves are difficult to reconcile with the recently reported data from $\upsilon$ Andromedae b \citep{Har06}, in which the phase offset between transit and the infrared minimum is much smaller and can be interpreted as arising from a situation in which very little heat is transferred from the day side to the night side.  \Citet{Har06} obtain a best fit with a phase shift of $11^{\circ}$, and the data are not inconsistent with zero phase shift.  We suggest two possible causes of this small phase shift.  First, at the 24 $\mu$m photosphere from which the data were taken, the pressure depth is quite low due to infrared absorption by water molecules.  Indeed, the 24-$\mu$m photosphere occurs at a pressure depth of 20 mbar \citep{For05}, at which $\tau_{rad} = 5.7$ hr \citep{Iro05}.  Our model predicts a smaller phase offset as $\tau_{rad}$ decreases.

In addition, our model indicates that lower atmospheric temperatures will also lead to smaller phase offsets.  Using planetary parameters appropriate to $\upsilon$ Andromedae b, an atmospheric temperature $T_{eff} = 1500$ K and a radiative timescale of 4 hr yields a phase offset of $49^{\circ}$; decreasing the atmospheric temperature to 800 K reduces the phase offset to $23^{\circ}$.  While more careful determination of the parameters and a more rigorous treatment of the radiative forcing will be necessary to obtain a good fit to the $\upsilon$ Andromedae b data, it seems quite plausible that the small phase offset is the result of a combination of the smaller radiative timescales and lower temperatures associated with the low pressure depths at which the measurements were made.

\section{Conclusions}

We have presented the results of a two-dimensional grid-based hydrodynamic simulation of the atmosphere of the transiting extrasolar giant planet HD 209458b.  In addition to the traditional case in which the planet's axial tilt is close to zero and the same side always faces the star, we have also considered the case that HD 209458b is tilted on its side, with an obliquity near $90^{\circ}$.  This configuration is dynamically stable, and is motivated by the need to explain the planet's anomalously large radius.

In both cases, a stable equilibrium develops, although the specific temperature distribution is markedly different.  In the zero-obliquity case, a steady state temperature profile is reached, characterized by a diffuse hot region blown around the globe by eastward jets at higher latitudes and a localized cold spot produced by a Bernoulli effect within an overall westward jet at the equator.  In the high-obliquity case, the temperature distribution is dominated by the contribution of radiative forcing, although the temperature extrema lead the forcing by $\sim 13$ hr.

In both cases, we expect to see observable changes in the planet's infrared spectrum; the maximum infrared emission is between 50\% (zero obliquity) and 70\% (high obliquity) greater than the minimum .  However, the difference in the light curves between the two states is likely to be difficult to observe experimentally, as both light curves are approximately sinusoidal.  In addition, the phase lag between illumination and infrared emission is similiar in both cases.

While the results of this preliminary investigation do not provide a clear observational distinction between the low- and high-obliquity cases, the results of \citet{Har06} demonstrate that analysis of the infrared emissions of extrasolar planets is a present possibility.  We believe that improvements to our model currently being tested will enable us to provide a good fit to the Harrington data and to other forthcoming observations of the infrared emissions of extrasolar planets.

\acknowledgements{We acknowledge support from the NASA Planetary Geology and Geophysics program through grant NNG04GK19G, and from the NSF through CAREER grant AST-0449986.  The code used for the simulations included the SPHEREPACK 3.0 subroutines provided by the University Corporation for Atmospheric Research.}

\clearpage

\clearpage

\clearpage

\begin{figure}
\begin{center}
\includegraphics[clip = true, bb=0in 0in 5in 6in,trim = 0in 0in 0  0in, width=4in]{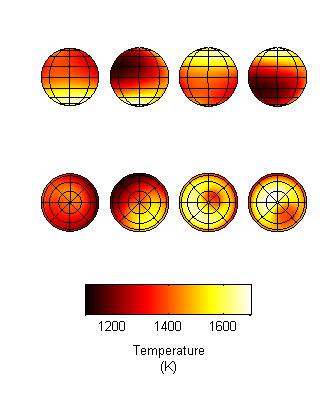}
\caption{Time-dependent temperature of HD 209458b in the high-obliquity case.  Each successive frame advances by one-fourth of a rotation period; the temperature variation is periodic, repeating once per rotation.  The top row shows an equatorial view, while a polar view can be seen on the bottom row.}
\label{CS2fig}
\end{center}
\end{figure}

\clearpage

\end{document}